\documentclass[final]{svjour3}
\usepackage{gensymb}
\usepackage{graphicx}
\usepackage{rotating}
\usepackage{amssymb}
\usepackage{mathptmx}
\usepackage{comment}
\usepackage{floatrow}
\usepackage[numbers]{natbib}
\makeatletter
\journalname{Journal of Low Temperature Physics}

\bibpunct{}{}{,}{s}{}{,}

\begin{document}

\newcommand{\hdblarrow}{H\makebox[0.9ex][l]{$\downdownarrows$}-}
\title{BICEP Array: 150 GHz detector module development
}

\author{A.~Schillaci$^1$ \and P.~A.~R.~Ade$^2$ \and Z.~Ahmed$^{3,4}$ \and M.~Amiri$^5$ \and D.~Barkats$^6$ \and R.~Basu Thakur$^1$ \and C.~A.~Bischoff$^7$ \and D. ~Beck$^{4}$ \and J.~J.~Bock$^{1,8}$ \and V.~Buza$^6$ \and J.~Cheshire$^9$ \and J.~Connors$^6$ \and J.~Cornelison$^6$ \and M.~Crumrine$^9$ \and A.~Cukierman$^4$ \and E. ~Denison$^{13}$\and M.~Dierickx$^6$ \and L.~Duband$^{10}$ \and M.~Eiben$^6$ \and S.~Fatigoni$^5$ \and J.~P.~Filippini$^{11,12}$ \and C.~Giannakopoulos$^7$ \and N.~Goeckner-Wald$^4$ \and D.~Goldfinger$^6$ \and J.~A.~Grayson$^4$ \and P.~Grimes$^6$ \and G.~Hall$^9$ \and G.~Halal$^4$ \and M.~Halpern$^5$ \and E.~Hand$^{7}$ \and S.~Harrison$^6$ \and S.~Henderson$^{3,4}$ \and S.~R.~Hildebrandt$^8$ \and G.~C.~Hilton$^{13}$ \and J.~Hubmayr$^{13}$ \and H.~Hui$^1$ \and K.~D.~Irwin$^{3,4}$ \and J.~Kang$^4$ \and K.~S.~Karkare$^{6,14}$ \and S.~Kefeli$^1$ \and J.~M.~Kovac$^6$ \and C.~L.~Kuo$^{3,4}$ \and K.~Lau$^9$ \and E.~M.~Leitch$^{14}$ \and A.~Lennox$^{11}$ \and K.~G.~Megerian$^8$ \and O.~Y.~Miller$^1$ \and L.~Minutolo$^1$ \and L.~Moncelsi$^1$ \and Y.~Nakato$^4$ \and T.~Namikawa$^{15}$ \and H.~T.~Nguyen$^8$ \and R.~O'Brient$^{8,1}$ \and S.~Palladino$^7$ \and M.~Petroff$^6$ \and  N.~Precup$^9$ \and T.~Prouve$^{10}$ \and C.~Pryke$^9$ \and B.~Racine$^6$ \and C.~D.~Reintsema$^{13}$ \and B. L.~Schmitt$^6$ \and B.~Singari$^9$ \and A.~Soliman$^1$ \and T.~St.~Germaine$^6$ \and B.~Steinbach$^1$ \and R.~V.~Sudiwala$^2$ \and K.~L.~Thompson$^{3,4}$ \and C.~Tucker$^2$ \and A.~D.~Turner$^8$ \and C.~Umilt\`{a}$^7$ \and C.~Verges$^6$ \and A.~G.~Vieregg$^{14}$ \and A.~Wandui$^1$ \and A.~C.~Weber$^8$ \and D.~V.~Wiebe$^5$ \and J.~Willmert$^9$ \and W.~L.~K.~Wu$^{14}$ \and E.~Yang$^4$ \and K.~W.~Yoon$^4$ \and E.~Young$^4$ \and C.~Yu$^4$ \and L.~Zeng$^6$ \and C.~Zhang$^1$ \and S.~Zhang$^1$}

\institute{$^1$Department of Physics, California Institute of Technology, Pasadena, California 91125, USA
\\$^2$School of Physics and Astronomy, Cardiff University, Cardiff, CF24 3AA, United Kingdom
\\$^3$Kavli Institute for Particle Astrophysics and Cosmology, SLAC National Accelerator Laboratory, 2575 Sand Hill Rd, Menlo Park, California 94025, USA
\\$^4$Department of Physics, Stanford University, Stanford, California 94305, USA
\\$^5$Department of Physics and Astronomy, University of British Columbia, Vancouver, British Columbia, V6T 1Z1, Canada
\\$^6$Harvard-Smithsonian Center for Astrophysics, Cambridge, Massachusetts 02138, USA
\\$^7$Department of Physics, University of Cincinnati, Cincinnati, Ohio 45221, USA
\\$^8$Jet Propulsion Laboratory, Pasadena, California 91109, USA
\\$^9$Minnesota Institute for Astrophysics, University of Minnesota, Minneapolis, 55455, USA
\\$^{10}$Service des Basses Temp\'{e}ratures, Commissariat \`{a} lEnergie Atomique, 38054 Grenoble, France
\\$^{11}$Department of Physics, University of Illinois at Urbana-Champaign, Urbana, Illinois 61801, USA
\\$^{12}$Department of Astronomy, University of Illinois at Urbana-Champaign, Urbana, Illinois 61801, USA
\\$^{13}$National Institute of Standards and Technology, Boulder, Colorado 80305, USA
\\$^{14}$Kavli Institute for Cosmological Physics, University of Chicago, Chicago, IL 60637, USA
\\$^{15}$Department of Applied Mathematics and Theoretical Physics, University of Cambridge, Wilberforce Road, Cambridge CB3 0WA, UK
\\$^{16}$Physics Department, Brookhaven National Laboratory, Upton, NY 11973
\\ \email{alex78@caltech.edu}}

\maketitle

\begin{abstract}

The BICEP/Keck Collaboration is currently leading the quest to the highest sensitivity measurements of the polarized CMB anisotropies on degree scale with a series of cryogenic telescopes, of which BICEP Array is the latest Stage-3 upgrade with a total of $\sim32,000$ detectors. The instrument comprises 4 receivers spanning 30 to 270 GHz, with the low-frequency 30/40 GHz deployed to the South Pole Station in late 2019. The full complement of receivers is forecast to set the most stringent constraints on the tensor to scalar ratio $r$. Building on these advances, the overarching small-aperture telescope concept is already being used as the reference for further Stage-4 experiment design.

In this paper I will present the development of the BICEP Array 150 GHz detector module and its fabrication requirements, with highlights on the high-density time division multiplexing (TDM) design of the cryogenic circuit boards. The low-impedance wiring required between the detectors and the first-stage SQUID amplifiers is crucial to maintain a stiff voltage bias on the detectors. A novel multi-layer FR4 Printed Circuit Board (PCB) with superconducting traces, capable of reading out up to 648 detectors, is presented along with its validation tests.

I will also describe an ultra-high density TDM detector module we developed for a CMB-S4-like experiment that allows up to 1,920 detectors to be read out. TDM has been chosen as the detector readout technology for the Cosmic Microwave Background Stage-4 (CMB-S4) experiment based on its proven low-noise performance, predictable costs and overall maturity of the architecture. The heritage for TDM is rooted in mm- and submm-wave experiments dating back 20 years and has since evolved to support a multiplexing factor of 64x in Stage-3 experiments. 


\end{abstract}

\section{Introduction}

The search for inflationary primordial gravitational waves in the CMB polarized signal is pushing the development of larger arrays of detectors. The Stage-3 class receivers currently being deployed in Antarctica and Chile are reaching the the current limits  the Time Division Multiplexing (TDM) readout capabilities. The recent outstanding results from the BICEP/Keck Collaboration [1, 2] show that improving further the sensitivity on the tensor to scalar ratio $r$, will require the development of multi-frequency $\sim10,000$ detector arrays. Synchrotron and dust polarized foregrounds as well as gravitational lensing distortion are presently the dominant factors on current $r$ constraints. 

 With this goal in mind the BK Collaboration is currently deploying the latest of the BK Series, the BICEP Array telescope [3]. This Stage-3 experiment comprises four $0.5$ m aperture telescopes working at $30/40$, $95$, $150$ and $220/270$ GHz. In the 2019-2020 austral summer the $30/40$ GHz low frequency receiver (BA1) [4] was successfully deployed at the Amundsen-Scott South Pole Station. The $\sim500$ detector array has already collected two years of data  aiming to place a more definite constraint on the faint Synchrotron polarized foreground in a $600$ deg$^{2}$ observing sky field. A refurbishment is taking place for the 2021-2022 austral summer with the goal of improving the BA1 receiver performances with new detector modules. In parallel the development of the three other BICEP-Array receivers is ongoing in North America with the plan of deploying the $150$ GHz receiver (BA2) for the 2022-2023 antarctic summer season. With close to $8000$ transition edge sensors (TES) [5] BA2 will have the highest density of detectors for the BK series ever. Section 2 of this paper will describe the general features of the receiver highlighting the innovative high density readout PCB that has been specifically developed  to achieve the wiring of this large number of detectors. 

With more than a decade of successfully deployed CMB polarization B-mode instruments, the BK Collaboration is playing a leading role within the CMB community. The recently formed CMB-S4 Collaboration [6]  is aiming to develop a large CMB observatory with more than a half million detectors to achieve extremely sensitive polarization measurements. The Small Aperture Telescope (SAT) baseline design is largely inspired on BK experience, where BICEP Array is regarded as the state of the art for this class of instruments. In section 3 a Detector Module Concept derived from BA2 experience will be presented. This concept design could potentially allow up to ~2000 detectors routing capability on a 6" hexagonal package.

\section{BICEP Array 150 GHz receiver}

The 150 GHz receiver design shares most of the hardware design with the other three receivers in BICEP Array [7]. We use a combination of a pulse tube refrigerator and a Helium-3/Helium-4 3-stages sorption fridge [8] to achieve a $~250$ mK FPU base temperature. The refractive f/1.55 2-lens HDPE optics is cooled to 4 K to reduce the radiative heat load on the detectors with the help of HDPE foam and alumina IR-blocker filters  at the 300 K and 50 K stages of the receiver. Thermo-mechanical structure and the magnetic shielding of the receiver have been simulated (in COMSOL/Solidworks) to minimize the impact of systematics in the CMB data and to ease data reduction.

\subsection{150 GHz receiver readout}

The 30/40, 95 and 150 GHz receivers all share the well-known and proven TDM readout architecture [9]. The last receiver of the series, BA4 at 220/270 GHz, is being developed using a frequency division multiplexing called $\mu$-MUX [10], due to the extremely large detector count . We also plan to include at least one module in the 220/270 GHz camera with thermal kinetic inductance detectors [11, 12, 13] as demonstration of an alternative technology that would greatly simplify hardware integration and reduce risk by reducing the wirebond count by orders of magnitude. Each receiver will be furnished with 12 detector modules each one carrying a 6" detector tile. The three TDM receivers have been designed to use the same modular readout package repeated enough times depending by the number of detectors to be read. 

The warm readout electronics are based on the Multi-Channel Electronics (MCE) developed by the University of British Columbia (UBC). Each MCE has 32 readout channels (columns) that can be switched between 41 addresses (rows) bringing the total count to $1312$ readable detectors per each of them.   

In table 1 we report the readout architecture numbers for the three TDM BICEP Array receivers. The BA2 receiver with 7776 Detectors will require 6 MCEs which presents an additional challenge of warm electronic space allocation. 

\begin{table}  
    \centering
    \begin{tabular}{|p{2cm}||p{3cm}|p{0.5cm}|p{3cm}|p{1.5cm}|}
  \hline
  Receiver observing band (GHz) &Detectors per module (per receiver) &Rows &Columns per module (per receiver)&MCEs per receiver\\
 \hline
  BICEP Array &&&& \\
$/$BA1 - 30 & 32 (192) & 33 & 2 (12) & 0.5  \\
$\setminus$BA1 - 40 & 50 (300) & 33 & 2 (12) & 0.5 \\
BA3 - 95 & 338 (4056) & 43 & 8 (96) & 3 \\
BA2 - 150 & 648 (7776) & 41 & 16 (192) & 6 \\

 \hline
 
\end{tabular}
    \caption{Readout channels count for the three TDM BICEP Array receivers at 30/40, 95 and 150 GHz. BA1 30/40 GHz receiver is split in 2 colors counting each as a 0.5 MCE. The 43 Rows for BA3 95 GHz Receiver will require an MCE firmware upgrade.}

    \label{tab:1}
\end{table}

\begin{figure}[htbp]
\begin{center}
\includegraphics[width=0.8\linewidth, keepaspectratio]{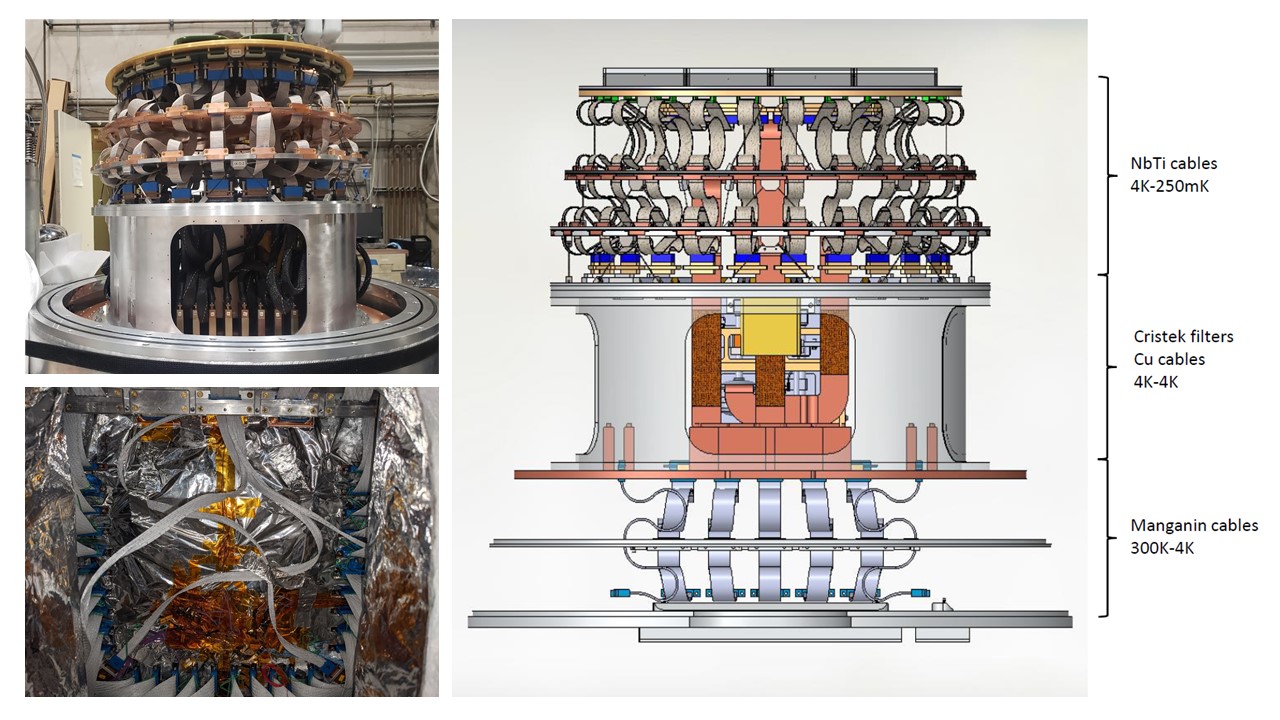}
\caption{\emph{Top/Left:} 150 GHz BA2 Insert with the NbTi cables and the 4 K isothermal cables and Cristek filters. \emph{Bottom/Left:} Bottom view of the 150 GHz BA2 300 K - 4 K manganine cables  \emph{Bottom/Right:} 3D Model showing the full readout cable chain installed (color figure online).}
\label{fig1}
\end{center}
\end{figure}

The readout uses $30 \times 100$ ways manganin cables, to carry electrical signals with minimal heat load from 300 K to the 50 K and 4 K stages of the cryostat (see fig\ref{fig1}). These low thermal conductivity cables are sized to minimize the conductive heat load from 300 K to the 50 K and 4 K stages of the cryostat. Entering the 4 K cryostat volume all wiring is passed through a stage of RF low pass filters \footnote{Cristek micro-D saver PI filtered, model no. C48-00063-01}. Next in the chain, simple copper iso-thermal cables connect to a stage of SQUID Series Array (SSA). A last set of $18$ $100$-ways Niobium-Titanium (NbTi) cables (3 per MCE) is used for the last section of the readout chain for connecting the SSA modules at 4 K up to a 0.250 K distribution board at the FPU. The choice of the NbTi Superconducting cable is driven by the low current signals produced by the  TES detectors which require a low-impedance circuitry. Also these cables have been carefully sized for preventing unwanted conductive heat load to reach the coldest volume of the instrument.

At the FPU, a large distribution PCB directs the signals to the 12 detector modules. The connection to this last step is made using 4 60-way cryogenic Kapton embedded copper Flat Flex Cable. 

\subsection{150 GHz module}

\begin{figure}[htbp]
\begin{center}
\includegraphics[width=0.8\linewidth, keepaspectratio]{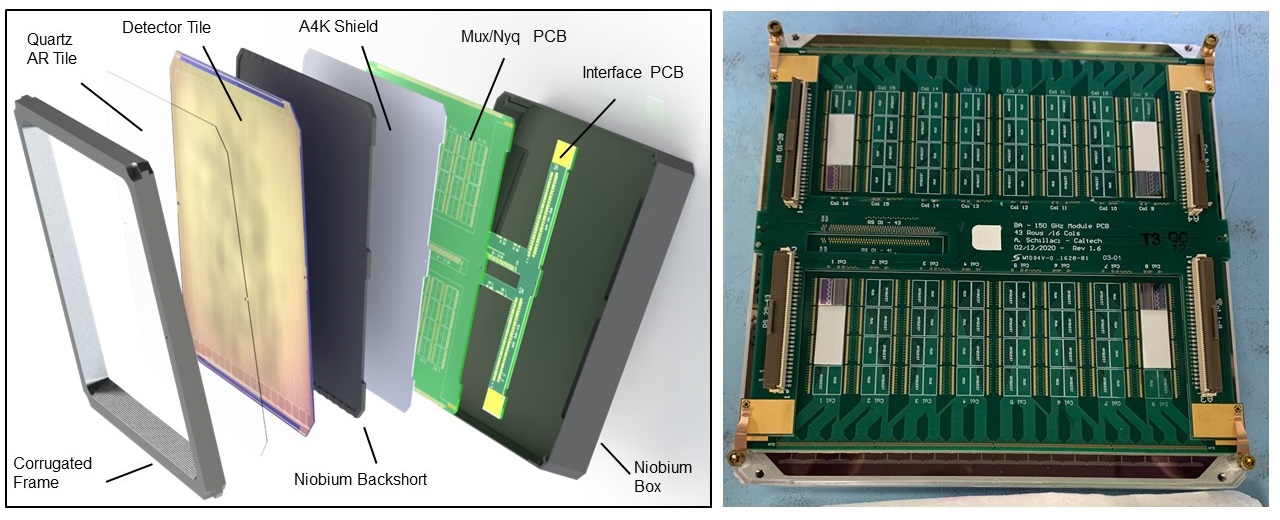}
\caption{\emph{Left:} Exploded view of the 150 GHz BA2 Detector Module. \emph{Right:} Open Test Module showing the module PCBs. Only 4 pairs of MUX/Nyq Chips are mounted (on ceramic carriers) for the purpose of the superconductivity test. The testing detector tile can be see at the top and bottom edges of the PCB. (color figure online).}
\label{fig2}
\end{center}
\end{figure}

The main architecture of the detector modules is shared among the 3 TDM receivers (Figure \ref{fig2}). The relevant opto-mechanical differences are driven by the different frequency assigned to a module. The frequency band of the receiver determines the design and optimization of the module $\lambda$/4 backshort, the quartz anti-reflection coating tile and the corrugated frames [14]. 
All of the BA detector modules have a high performance magnetic shield in the form of a niobium box plus a inner sheet of A4K high magnetic permeability steel for protecting the multiplexing chips carrying the SQUIDS (for a detailed description of the magnetic shielding see [3]). These are NIST Nyquist and MUX chips [15, 16] and they are glued and wirebonded on aluminum nitride substrate with aluminum traces single layer PCBs. This PCB provides the routing to interface with the detector tile above it.  The choice of superconductive aluminum traces is driven by the necessity to have low resistance circuitry between the detector tile and the MUX chips. 

This choice worked flawlessly with the low frequency receiver BA1 but the large number of traces that will need to be connected for BA2, required the development of a new FR4 multi-layer PCB. We found trying to achieve this density with a commercial single layer PCB was costly and low yield. 

Comparing the commercial prices for aluminum nitride single layer and multi-layer FR4 PCBs, the former can cost up to $\$5,000$ per each board while the latter being about 20 times cheaper ($\$250$ each). 

\subsection{150 GHz MUX/Nyq interface PCB superconductivity test}

\begin{figure}[htbp]
\begin{center}
\includegraphics[width=0.8\linewidth, keepaspectratio]{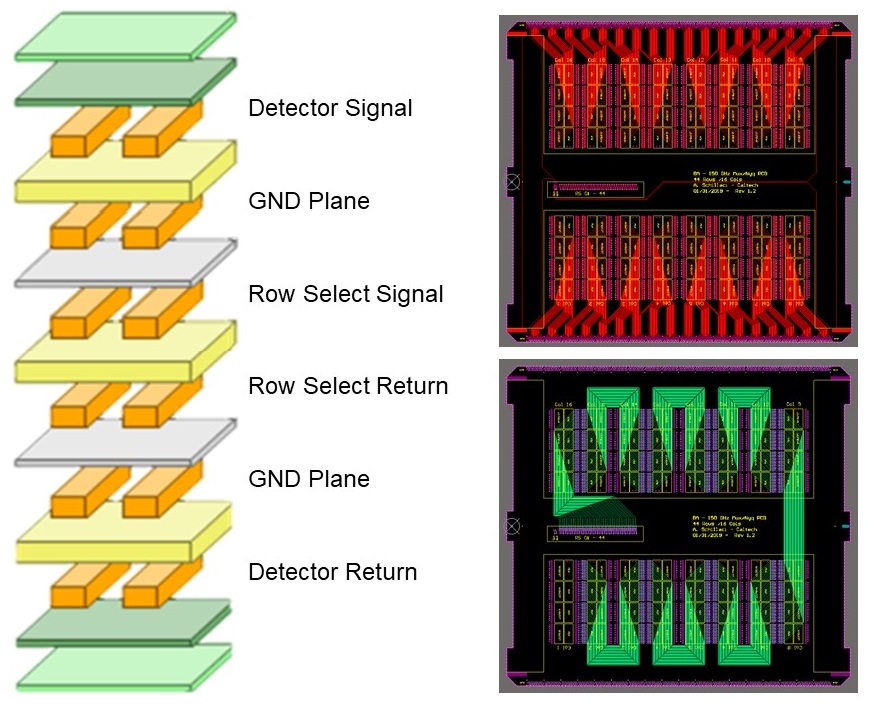}
\caption{\emph{Left:} Exploded view of the layer stack for the 150 GHz MUX/Nyq PCB. The top and bottom Layers are the ones that go to the detectors and that need the tin/lead plating \emph{Right/Top:} 150 GHz MUX/Nyq PCB top layer carrying the detector signals (Detector Returns are similarly routed on the bottom layer)  \emph{Right/Bottom:} 150 GHz MUX/Nyq PCB middle layer carrying the row select signals (the row select returns are similarly routed on a second middle layer). (color figure online).}

\label{fig3}
\end{center}
\end{figure}

Referring to Figure \ref{fig3}, the 150 GHz MUX/Nyq PCB is a compact and high trace density  6-layers FR4 design. Traces are have a copper weight of $0.5$ oz (equivalent to $0.0175$ mm thickness) with a width/pitch of $3$ / $6$ mil ($76$ / $152$  micron). Two middle layers have been used for the row select routing and have been packed in between 2 ground planes in order to minimize the cross-talk. The outer top and bottom layers have been used for the detector wiring that requires superconductive traces. To achieve that, we asked the manufacturer (TTM Technologies, North Jackson \footnote{https://www.ttm.com/}) to plate the copper traces with a tin/lead unfused plating with a minimum thickness of 100 microinches (2.54 microns). After manufacturing the achieved thickness has been measured by TTM to have an average value of 400 microinches, that is comfortably above our minimum requirement. 

\begin{figure}[htbp]
\begin{center}
\includegraphics[width=0.8\linewidth, keepaspectratio]{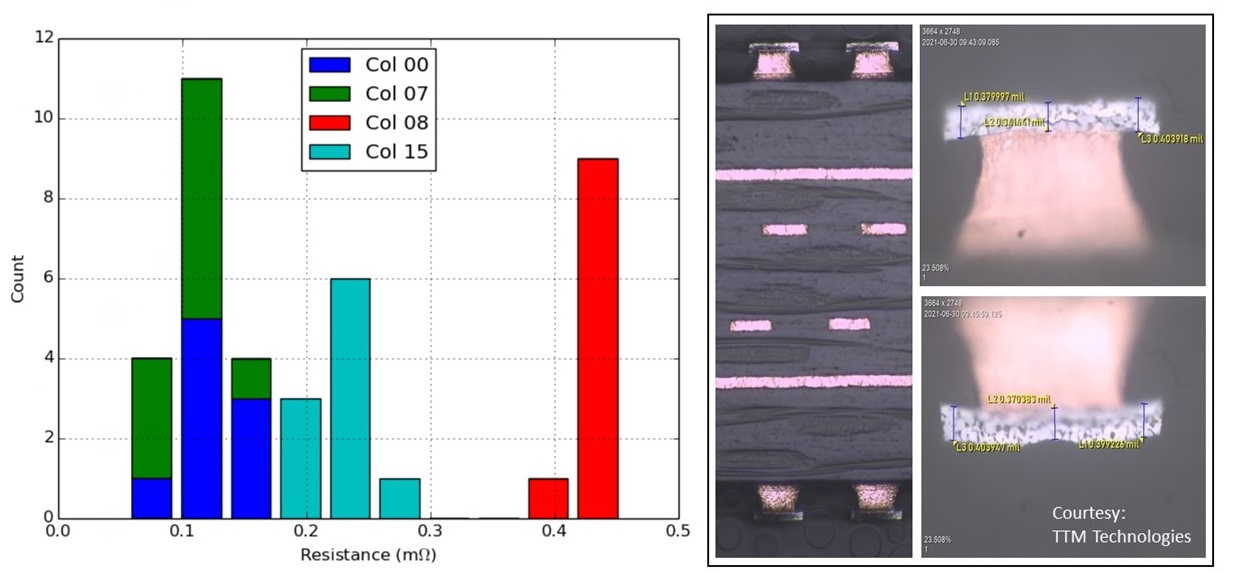}
\caption{\emph{Left:} Resistance measurements histogram on the 40 channels we tested in the MUX/Nyq PCB.  \emph{Right:} Pictures of one sliced 150 GHz MUX/Nyq PCB showing the layer stack. In the zoom panels the Top and Bottom traces Sn/Pb unfused plating is shown and its thickness measured at about 400 microinches. Pictures courtesy TTM Technologies, North Jackson. (color figure online).}
\label{fig4}
\end{center}
\end{figure}

Before assembling a fully chip-populated  150 GHz module (each unit requires $64\times$ MUX and 64 Nyquist Chips), we performed a superconductivity test on the top and bottom Sn/Pb plated traces of this PCB. This test unit with just 4 chip pairs is shown in the right panel of Figure \ref{fig2}. It was tested in a $0.25$ K test-bed cryostat at Caltech. We tested a sample of 40 channels within the 648 available. The channels were distributed in 4 groups of 10, one group per each quadrant of the PCB in order to monitor the uniformity of the Sn/Pb Plating. The channels are terminated with a niobium  superconducting short deposited on the test silicon wafer in  place of the TES detector. In this way we can avoid any additional in series contribution that could affect our measurements. The plot on the left panel of figure \ref{fig4} shows the electric current in the traces versus the input bias electric current for one of the 10-Channels sample, showing a very low residual resistance and verifying the achieved superconductivity of the Traces. 

Since this PCB design has been proven to work, we are currently working on assembling the first 150 GHz detector module. Our plan is to assemble and test 12 units for the scheduled BA2 receiver deployment in the 2022/2023 austral summer.

\section{CMB Stage-4 high density Detector Module Concept}

CMB Stage-4 is meant to deploy $\sim500,000$ TES Detectors in the next decade between South Pole and the Atacama Desert. These large number of new telescopes will range between 20 GHz up to 270 GHz. The highest frequency modules will approach 2,000 TES Detectors on an hexagonal 6" tile, making the design of a MUX/Nyq PCB very challenging. 

\begin{figure}[htbp]
\begin{center}
\includegraphics[width=0.8\linewidth, keepaspectratio]{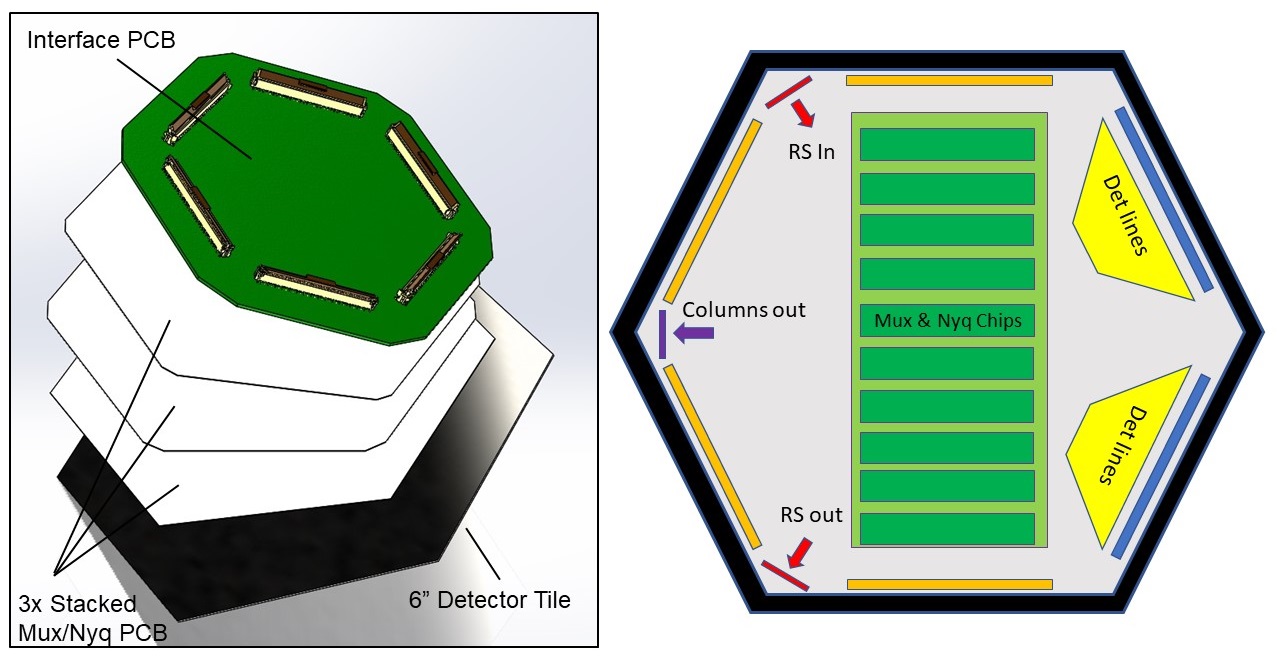}
\caption{\emph{Left:} 3D exploded model of the staked multi-layer PCB design. The interface PCB and the 6" detector tile are shown in the stack. \emph{Right:} Concept diagram of the single MUX/Nyq PCB showing the location of the wirebond pad groups at the edges. Each PCB will serve two sides of the hexagon. Also the column and row select wirebond pad groups buses that will route between the 3 PCBs are shown. (color figure online).}
\label{fig5}
\end{center}
\end{figure}

In figure \ref{fig5} a readout concept design with 3 multi-layer MUX/Nyq PCB stacked and interconnecting wirebond buses is illustrated. With 64 rows and 30 columns distributed within the 3 stacked PCBs, the design could potentially achieve the required large wiring capability of the next generation Stage-4 high frequency receiver modules. 

On this framework the experience with the development of the BICEP Array MUX/Nyq PCB described in this paper, can be a valuable reference for the next generation TDM instrument readout.

\begin{acknowledgements}
The BICEP/Keck project has been made possible through a series of grants from the National Science Foundation including 0742818, 0742592, 1044978, 1110087, 1145172, 1145143, 1145248, 1639040, 1638957, 1638978, 1638970, \& 1726917 and by the Keck Foundation.The development of antenna-coupled detector technology was supported by the JPL Research and Technology Development Fund and NASA Grants 06-ARPA206-0040, 10-SAT10-0017, 12-SAT12-0031, 14-SAT14-0009 \& 16-SAT16-0002. The development and testing of focal planes were supported by the Gordon and Betty Moore Foundation at Caltech.Readout electronics were supported by a Canada Foundation for Innovation grant to UBC.The computations in this paper were run on the Odyssey cluster supported by the FAS Science Division Research Computing Group at Harvard University. The analysis effort at Stanford and SLAC is partially supported by the U.S. DoE Office of Science.We thank the staff of the U.S. Antarctic Program and in particular the South Pole Station without whose help this research would not have been possible.Tireless administrative support was provided by Kathy Deniston, Sheri Stoll, Nancy Roth-Rappard, Irene Coyle, Donna Hernandez, and Dana Volponi. 

\end{acknowledgements}

\pagebreak

\end{document}